\documentclass[fleqn,10pt]{wlscirep}
\usepackage[utf8]{inputenc}
\usepackage[T1]{fontenc}
\usepackage{graphicx}
\usepackage{epsfig}
\usepackage{epstopdf}
\title{Order-disorder phase transition and elastic-to-plastic vortex creep crossover in a triclinic iron pnictide superconductor (Ca$_{0.85}$La$_{0.15}$)$_{10}$(Pt$_3$As$_8$)(Fe$_2$As$_2$)$_5$}

\author[1,2]{Shyam Sundar}
\author[1]{P. V. Lopes}
\author[1]{S. Salem-Sugui, Jr.}
\author[3]{Z. -Z. Li}
\author[3,4]{W. -S. Hong}
\author[3]{H. -Q. Luo}
\author[3]{S. -L. Li}
\author[1]{L. Ghivelder}
\affil[1]{Instituto de Fisica, Universidade Federal do Rio de Janeiro, 21941-972 Rio de Janeiro, RJ, Brazil}
\affil[2]{School of Physics and Astronomy, University of St Andrews, St Andrews KY16 9SS, Scotland, United Kingdom.}
\affil[3]{Beijing National Laboratory for Condensed Matter Physics, Institute of Physics, Chinese Academy of Sciences, Beijing 100190, China}
\affil[4]{International Center for Quantum Materials, School of Physics, Peking University, Beijing 100871, China}

\begin{abstract}
Vortex matter in layered high-$T_c$ superconductors, including iron-pnictides, undergo several thermodynamic phase transitions due to the complex interplay of pinning energy, thermal energy and elastic energy. Moreover, the presence of anisotropy makes their vortex physics even more intriguing. Here, we report a detailed vortex dynamics study, using dc magnetization measurements, in a triclinic iron-pnictide superconductor (Ca$_{0.85}$La$_{0.15}$)$_{10}$(Pt$_3$As$_8$)(Fe$_2$As$_2$)$_5$, with a superconducting transition temperature, T$_c$ $\sim$ 31 K. A second magnetization peak (SMP) feature is observed for magnetic field perpendicular ($H$$\parallel$$c$) and parallel ($H$$\parallel$$ab$) to the crystal plane. However, its fundamental origin is quite different in both directions. For $H$$\parallel$$c$, the SMP can be well explained using an elastic-to-plastic vortex creep crossover, using collective creep theory. In addition, a possible rhombic-to-square vortex lattice phase transition is also observed for fields in between the onset-field and peak-field related to the SMP. On the other hand, for $H$$\parallel$$ab$, a clear signature of an order-disorder vortex phase transition is observed in the isothermal $M$($H$) measurements at $T$ $\geq$ 6 K. The disordered phase exhibits the characteristics of entangled pinned vortex-liquid. We construct a comprehensive vortex phase diagram by displaying characteristic temperatures and magnetic fields for both crystal geometries in this unique superconducting compound. Our study sheds light on the intricate vortex dynamics and pinning in an iron-pnictide superconductor with triclinic symmetry.
\end{abstract}
\begin{document}

\flushbottom
\maketitle
\thispagestyle{empty}

\section*{Introduction}

Vortex-matter state in high temperature superconductors, including iron-pnictides, copper oxides, and iron chalcogenides, consists of various interesting phases due to their layered crystal structure, low coherence length ($\xi$) and high superconducting transition temperature ($T_c$) \cite{Kwok:2016, Blatter1994}. Second magnetization peak (SMP) is one of such captivating feature, which appears in the isothermal magnetization curve, $M$($H$), measured in the superconducting state. Although, the SMP is common in both conventional \cite{Lortz:2007, Stamopoulos:2004} and in unconventional superconductors \cite{Sundar:2019a, Said:2020, Eley:2020}, its origin has been found to be fundamentally different and sample dependent \cite{Wang:2021, Miu2020}. Therefore, there has been a continuous effort to understand the origin of this intriguing vortex-matter phase in different superconducting materials, especially in high-$T_c$ cuprates \cite{Said:2020, Eley:2020} and iron-pnictide superconductors \cite{Llovo:2021}. A wide variety of mechanisms has been reported to be responsible for the SMP in different superconducting materials, such as, elastic-plastic pinning crossover \cite{Sundar2017, SalemSugui2010}, order-disorder vortex phase transition \cite{Miu2012, Hecher2014, Zehetmayer2015}, vortex lattice melting \cite{Kwok1994, Tamegai1996}, weak to strong pinning crossover \cite{Galluzzi2018} and also a rhombic-square vortex lattice phase transition \cite{Rosenstein2005}. Most of the vortex dynamics studies on iron-pnictides have been performed on materials with tetragonal crystal structure and low anisotropy. A vortex dynamics study in a more anisotropic, less symmetric iron-pnictide superconductor would be interesting to explore different vortex matter phases.

The cuprates are categorized and widely known as anisotropic superconductors due to their layered structure containing CuO$_2$ planes. Above a characteristic lower critical field, $H_{c1}$, magnetic field applied perpendicular to the CuO$_2$ planes penetrates in the form of 2-dimensional (2D) vortex lattice within the CuO$_2$ planes. However, the 2D nature of vortices depends on the strength of the Josephson coupling between the layers. A three-dimensional (3D) vortex lattice in the form of continuous string of magnetic field is often observed with strong Josephson coupling between CuO$_2$ planes. The 2D vortices or pancake vortices, are pinned individually within the CuO$_2$ plane, however, a 3D vortex may interact with many pinning centers. On the other hand, sufficiently high magnetic field applied parallel to the CuO$_2$ planes penetrates in the form of the Josephson vortices within the space between planes. Contrary to the 3D vortices, these Josephson vortices do not have normal core and are  localized between the CuO$_2$ planes due to the intrinsic pinning. For the details on vortices in layered high-$T_c$ superconductors referred review articles would be interesting to follow \cite{Kwok:2016, Blatter1994}. As an effect of different pinning landscapes, combined with thermal fluctuations and vortex elastic energy, the magnetic field applied parallel and perpendicular to the crystal plane may lead to the different vortex dynamics in highly anisotropic layered superconductors.

Soon after the breakthrough of superconductivity in iron-pnictides in 2008, superconductivity was realized in two Pt containing iron-arsenide compounds, namely Ca$_{10}$(Pt$_3$As$_8$)(Fe$_2$As$_2$)$_5$ (the "10-3-8 phase") and Ca$_{10}$(Pt$_4$As$_8$)(Fe$_2$As$_2$)$_5$ (the "10-4-8 phase") \cite{Kakiya2011, Ni2011}. However, only the "10-3-8 phase" achieves superconductivity via electron doping at the Fe site with Pt atoms. Subsequently, it was observed that partial substitution of Ca with La significantly enhances superconductivity in the 10-3-8 phase compared to replacing Fe with Pt \cite{Stuerzer2012}. The 10-3-8 phase possess a less symmetric triclinic crystal structure which is quite rare in superconductors, whereas, likewise in other iron-pnictides, the 10-4-8 phase possess a high symmetry tetragonal structure. Most of the common features of iron-pnictide superconductors are also observed in the 10-3-8 and 10-4-8 phases, such as, multiband superconductivity \cite{Seo2017}, large upper critical field, $H_{c2}$ \cite{Ni2013}, and high critical current density, $J_c$ \cite{Choi2020}. However, in contrast to other iron-pnictide superconductors, 10-3-8 and 10-4-8 phases have intermediary Pt$_n$As$_8$ ($n$ = 3, 4) layers, that leads to the Ca-(Pt$_n$As$_8$)-Ca-Fe$_2$As$_2$-layer stacking. Here, the coupling between the FeAs layers is controlled through the intermediary layers of PtAs which in turn are responsible for the $T_c$ enhancement. This structure is similar to the one in high-$T_c$ Bi$_2$Sr$_2$Ca$_{n-1}$Cu$_n$O$_{2n+4-x}$ (BSCCO, $n$ =1-3) systems, where the high-$T_c$ is controlled by the coupling between the CuO$_2$ planes through the intermediary layers. Furthermore, similar to cuprates, a relatively high anisotropy near $T_c$, $\gamma$ $>$ 10, high Ginzburg number, $G_i$ $\sim$ 0.1 and small superfluid density are also observed in the 10-3-8 phase \cite{Xiang2012, Surmach2015, Pan2017}. Such high anisotropy was also observed in another iron pnictide KCa$_2$Fe$_4$As$_4$F$_2$, which has a bilayer structure and a possible pseudogap behavior similar to BSCCO \cite{Wang2020, Zhang2022, Hao2022}. Moreover, interesting vortex dynamics behaviour has also been observed in a recent study on anisotropic KCa$_2$Fe$_4$As$_4$F$_2$ compound \cite{Lopes2022}.  These salient features, in addition to the triclinic crystal structure (contrary to the tetragonal crystal structure in cuprates and other iron pnictide superconductors), make the 10-3-8 phase a unique compound to explore interesting and possibly new vortex phases. 

In this paper, we investigated the vortex dynamics and pining behaviour in a single crystal of triclinic iron pnictide superconductor (Ca$_{0.85}$La$_{0.15}$)$_{10}$(Pt$_3$As$_8$)(Fe$_2$As$_2$)$_5$ (hereafter referred to as "La-10-3-8"), with $T_c$ $\sim$ 31 K, via detailed measurements of DC magnetization as a function of temperature, magnetic field and time. Experiments were performed for magnetic field direction in-plane, $H$$\parallel$$ab$, as well as, out-of-plane, $H$$\parallel$$c$, of the crystal. The SMP features were observed in the isothermal $M$($H$) measurements for both directions. In contrast to the SMP observed for $H$$\parallel$$c$ in the entire temperature range below $T_c$, the $H$$\parallel$$ab$ measurements show two distinct SMP features at low-$T$ and high-$T$ regions below $T_c$. A detailed analysis of magnetic relaxation data obtained for $H$$\parallel$$c$ evidenced a structural phase transition just before the SMP, which is followed by an elastic to plastic pinning crossover across the SMP. On the other hand, for $H$$\parallel$$ab$, the SMP is realized due to the order-disorder phase phase transition. These claims were substantiated with a detailed analysis of magnetic relaxation data using the collective pining theory. Moreover, a marked change of slope in $M$($H$) below SMP is noticed as the signature for such order-disorder phase transition. Interesting $H$-$T$ phase diagrams for both directions are presented. Furthermore, pinning behaviour in both directions is also explored by analysing the temperature dependence of critical current density and magnetic field dependence of pinning force curves. 

\section*{Results \& Discussion}

\begin{figure}[htb]
\includegraphics[width=\linewidth]{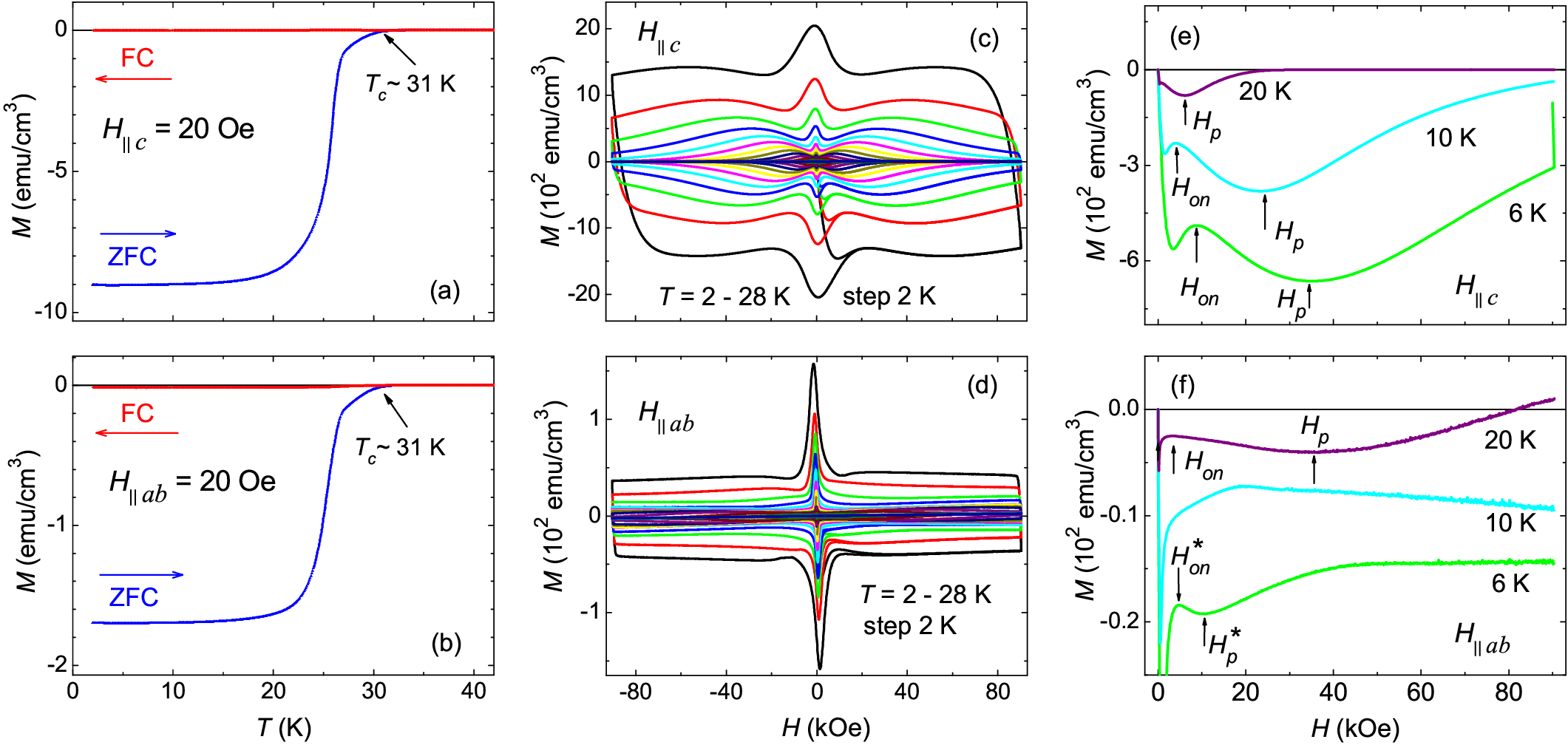}
\caption{Temperature dependence of magnetization, $M$($T$), in zero-field cooled (ZFC) and field-cooled (FC) states measured with a 20 Oe magnetic field applied parallel (a), and perpendicular (b) to the crystal plane. The onset of the diamagnetic transition, defined as the superconducting transition temperature $T_c$, is shown by an arrow in both panels. (c, d) Isothermal magnetic field dependence of the magnetization, $M$($H$), measured in a temperature range 2-28 K for $H$$\parallel$$c$, and $H$$\parallel$$ab$ respectively. Both geometries show a second magnetization peak (SMP) in their respective isothermal $M$($H$) curves. (e) Initial branch of the isothermal $M$($H$) curves at 6 K, 10 K, and 20 K for $H$$\parallel$$c$ showing the usual decrease of onset-field ($H_{on}$), and peak-field ($H_p$), with increasing temperature. (f) On the contrary, for $H$$\parallel$$ab$, the $H_{on}$ and $H_p$ values for 6 K are lower than the ones for 10 K and 20 K. This suggests the different nature of SMPs observed at low temperatures and at high temperatures for $H$$\parallel$$ab$.}
\label{fig:fig1}
\end{figure}

Figures 1(a) and 1(b) show the temperature dependence of magnetization in ZFC and FC modes measured with a 20 Oe magnetic field for $H$$\parallel$$c$ and $H$$\parallel$$ab$ respectively. For both orientations, a clear diamagnetic transition in ZFC magnetization is observed at $\sim$31 K, which is defined as the $T_c$ of the sample. Moreover, the small change in FC magnetization below $T_c$ (small Meissner fraction) indicates the strong pinning for both crystal orientations. Figure 1(c) shows the ZFC isothermal magnetic hysteresis, $M$($H$), for $H$$\parallel$$c$ in the temperature range 2-28 K. A well defined SMP feature is observed in the entire temperature range below $T_c$. The characteristic peak field, $H_p$, and onset field, $H_{on}$, associated with the SMP are clearly marked in Fig. 1(e), where, $H_p$ and $H_{on}$ decrease as the temperature approaches towards $T_c$. Contrary to the $H$$\parallel$$c$, the isothermal $M$($H$) for $H$$\parallel$$ab$ in Fig. 1(d) shows two distinct SMP features appeared in $M$($H$) measured at low-$T$ and high-$T$. This is more explicitly demonstrated in Fig. 1(f), where, the peak field at 20 K is found to be higher than the peak field at 6 K. Moreover, the observed shape of the $M$($H$) near onset-field is quite different at 10 K than the one at 6 K. This clearly shows the distinct behaviour of SMP at low-$T$ and at high-$T$ for $H$$\parallel$$ab$. In order to distinguish the effects, for low-$T$ SMP, the peak field and onset-field are labeled as $H^*_p$, and $H^*_{on}$ in Fig. 1(f). 

Figures 2(a) and 2(c) show the initial branch of isothermal $M$($H$) for $H$$\parallel$$ab$, measured at 20 K and 10 K respectively. The expanded view of part of the data, shown in Fig. 2(b) and 2(d), exhibit a "kink feature", at fields $H_k$, just above the onset field, $H_{on}$. While $H_k$ can be readily identified in $M$($H$) curves at $T$ < 20 K, its determination becomes challenging at $T$ $\geq$ 20 K (see Fig. 2(b)), owing to the broader manifestation of the kink feature at elevated temperatures. In addition to a SMP for fields below 15 kOe (Fig. 1(f)), a weak onset of SMP-like feature is also observed at higher fields in $M$($H$) at $T$ = 6 K, shown in Fig. 2(e) (dashed rectangle). The expanded view of Fig. 2(e) at higher fields is shown in Fig. 2(f) where the onset field associated to the weak SMP and a kink feature ($H_k$) similar to one observed at higher temperatures ($T$$>$6 K) are also present. In the literature, a similar kink feature just above the onset peak associated to the SMP, has been observed in cuprate superconductors \cite{Pissas2000, Giller1997}. In cuprates, the kink feature, $H_k$, is linked to a transition from a relatively ordered vortex lattice to the highly disordered entangled vortex solid phase. It is theoretically predicted and experimentally verified in cuprates that the disordered vortex solid phase above $H_k$ also leads to the plastic vortex creep way before approaching $H_p$ \cite{Giller1997, Ertas1996, Giamarchi1997}. Moreover, it is worth noticing that just above $H_k$ a pronounced noise in the $M$($H$) data is also observed. Contrary to the $H$$\parallel$$ab$ case, we did not observe a kink feature in $M$($H$) for $H$$\parallel$$c$. This absence of the kink feature has also been noted in LSCO and YBCO superconductors, where a kink feature in $M$($H$) is evident for $H$$\parallel$$ab$, but not for $H$$\parallel$$c$ \cite{Giller1999, Radzyner2002}. In the cases of LSCO and YBCO superconductors, it has been attributed to the presence of twin boundaries within the crystal lattice. Therefore, in addition to anisotropy, it is likely that the specific defects in our sample, akin to twin boundaries, might be responsible for the missing kink feature in $M$($H$) for $H$$\parallel$$c$. Although twin-boundaries are common in iron-pnictide superconductors \cite{Tanatar2009, Kalisky2011}, we currently lack information regarding the existence of twin-boundary defects in our crystal. In the latter part of the paper, we will address the details of the vortex dynamics across the SMPs observed for $H$$\parallel$$c$ as well as for $H$$\parallel$$ab$. In addition, we will demonstrate that the noise in $M$($H$) data above $H_k$ is related to the vortex dynamics in the disordered vortex phase in the sample.    

\begin{figure}[htb]
\includegraphics[width=\linewidth]{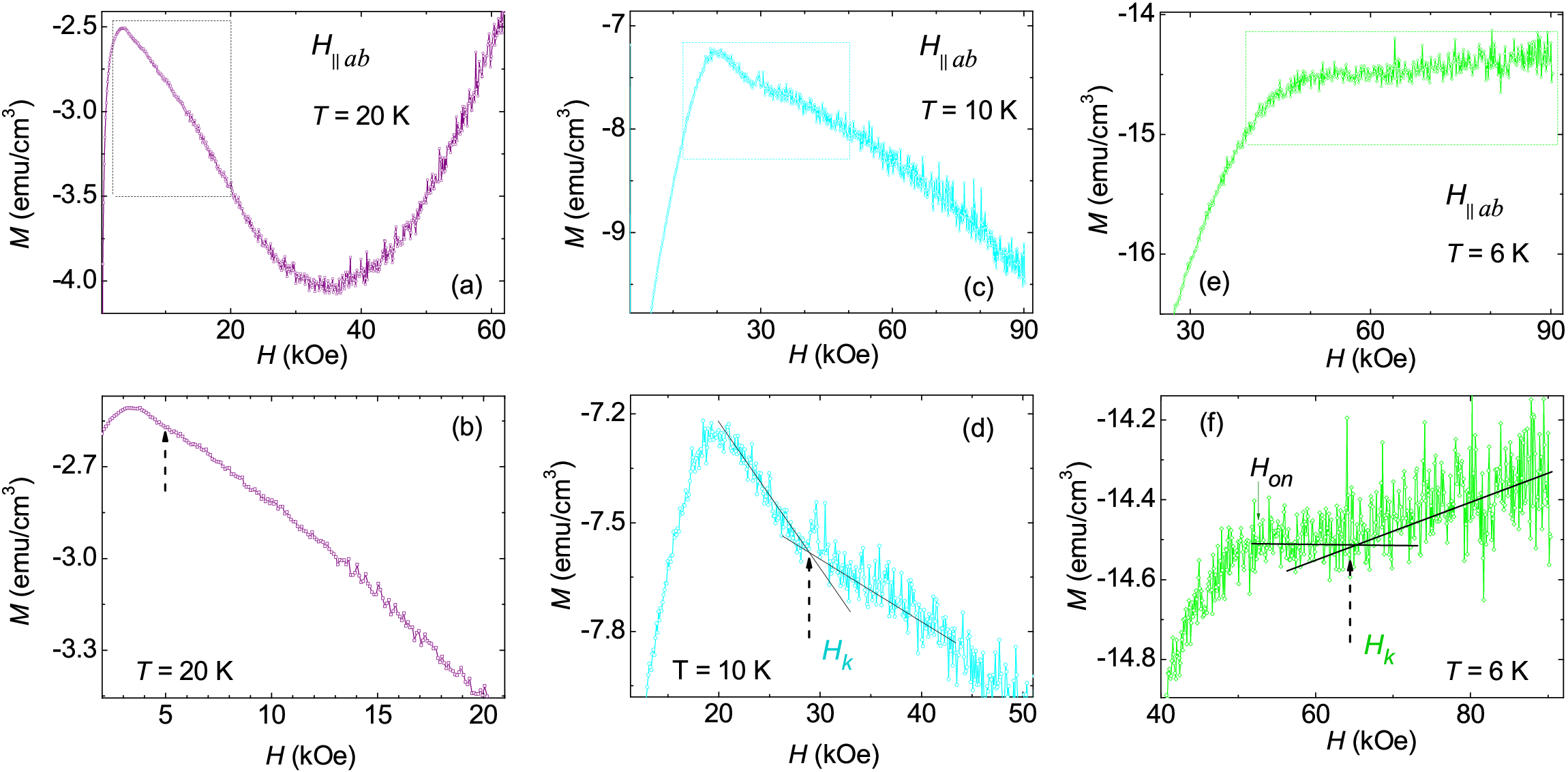}
%\begin{center}
%\includegraphics[scale=0.45]{Fig2.eps}
%\end{center}
\caption{Initial branch of the isothermal magnetic field dependence of magnetization measured for $H$$\parallel$$ab$ at $T$ = (a) 20 K, (c) 10 K, and (e) 6 K. The enlarged view of the dotted rectangular region is shown in panels (b), (d), and (f). A kink feature, $H_k$, reminiscent of the order-disorder phase transition, is clearly observed just above $H_{on}$. Moreover, a pronounced noise in isothermal $M$($H$) is noticed for fields above $H_k$.}
\label{fig:fig2}
\end{figure}

Magnetic relaxation measurements were performed on the initial branch of the isothermal $M$($H$) at several fixed magnetic fields. A representative magnetic relaxation data, on the initial branch of isothermal $M$($H$), measured at 15 K for $H$$\parallel$$c$ is shown in the inset (i) of Fig. 3. The dotted line, a guide to the eyes, exhibit the variation of the final magnetization with magnetic field after relaxation, which follows the same magnetic field dependence as in the initial $M$($H$) branch. Magnetic relaxation data for $H$$\parallel$$c$ follows a typical logarithmic behaviour, ln$\lvert M \rvert$ $\sim$ ln($t$). This allowed us to extract the relaxation rate, $R$ = dln$\lvert M \rvert$/dln$t$, by plotting ln$\lvert M \rvert$ as a function of ln$t$. Inset (ii) of Fig. 3 displays the straight lines fits to the ln$\lvert M \rvert$ $vs.$ ln$t$ curves measured at 15 K for the entire field range $H_p<H<H_{on}$. Similar relaxation measurements were performed on the various isothermal $M$($H$) below $T_c$ for $H$$\parallel$$c$, and the obtained relaxation rate as a function of magnetic field, $R$($H$), is plotted in Fig. 3 for different temperatures. Each $R$($H$) for $H$$\parallel$$c$ shows a well defined dip, $H_{cr}$, at intermediate magnetic fields for each temperature. Such dip feature in isothermal $R$($H$) has been observed previously in Co and Ni-doped "122" iron-pnictide superconductors \cite{Sundar:2019a, Sundar2017a, Kopeliansky2010}, and has been interpreted in terms of a rhombic to square vortex lattice phase transition. However, this also requires a similar dip feature in the isofield temperature dependence of relaxation rate, $R$($T$), which is discussed later in the paper.   

\begin{figure}[htb]
\begin{center}
\includegraphics[scale=0.5]{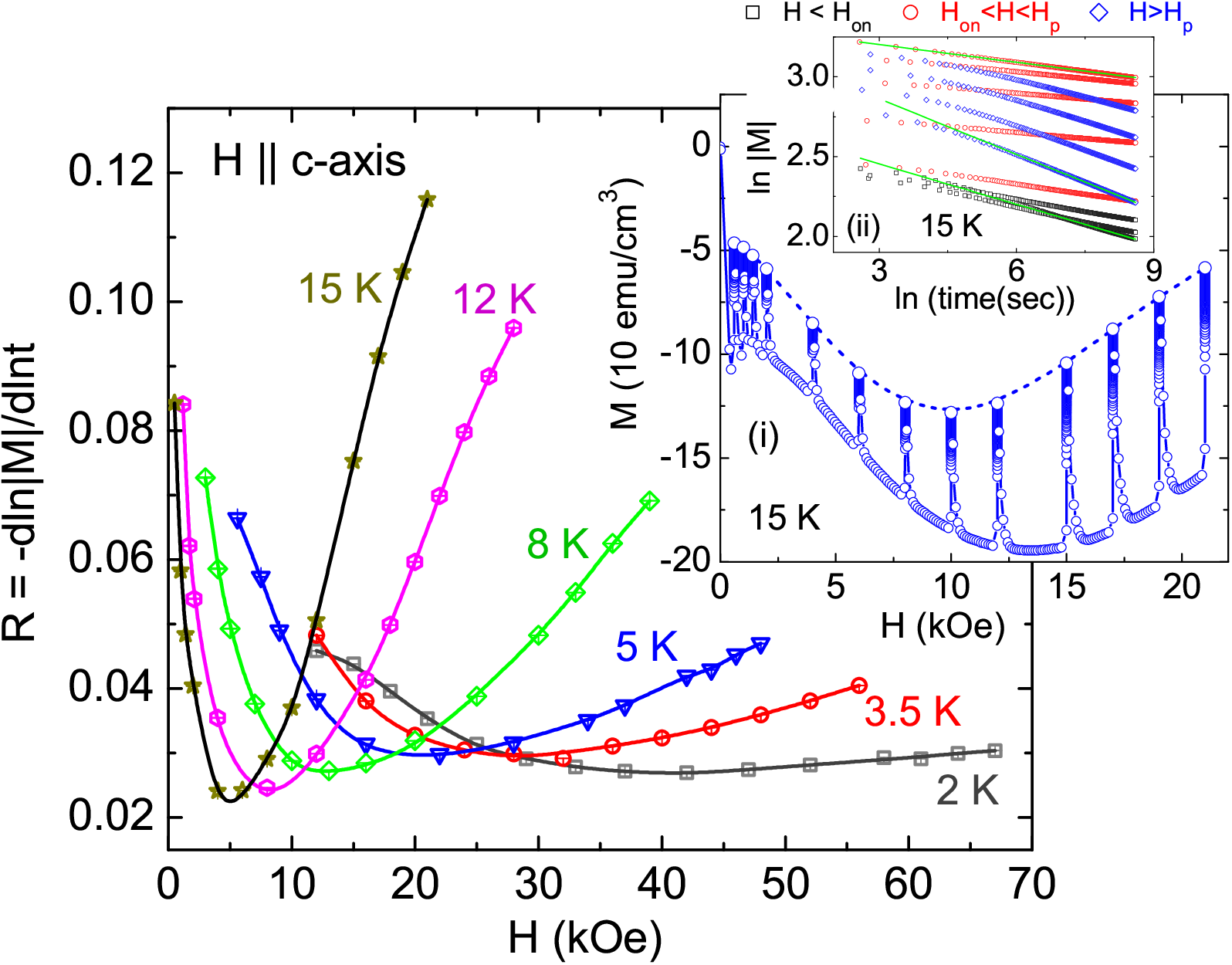}
\end{center}
\caption{Magnetic relaxation rate, $R$ = dln$\lvert M \rvert$/dln$t$, as a function of magnetic field at fixed temperatures for $H$$\parallel$$c$ (solid lines are guide to eyes). A dip, marked as $H_{cr}$, is clearly observed in each isothermal $R$($H$) curve, where, $H_{cr}$ increases as temperature decreases. Inset (i) shows the magnetic relaxation measured at the initial branch of the $M$($H$) at 15 K for fixed magnetic fields ranging from $H$ = $H_{on}$ to $H$$>$$H_p$. Interestingly, it is noticed that the magnetic relaxation suppress the peak-field, $H_p$ (see the dotted line in the inset (i)). Inset (ii) shows the linear behaviour of ln$\lvert M \rvert$ vs. ln($t$) curves obtained for 15 K.}
\label{fig:RH_c-axis}
\end{figure}

Figure 4(a) shows the initial branch of the isothermal $M$($H$) obtained at 18 K for $H$$\parallel$$ab$. A kink feature is marked as $H_k$, followed by a SMP at higher fields. In addition, a pronounced noise is observed for $H$$>$$H_k$. Similar observation has also been noticed at other temperatures above 6 K. Magnetic relaxation measurements have been performed for $H$$\parallel$$ab$ to understand the origin of the kink and the pronounced noise in the isothermal $M$($H$) curves. Relaxation measurements on the initial branch of $M$($H$), measured at 18 K is represented in the inset (i) of Fig. 4(a). Here, it is interesting to note that at low fields ($H$$<$10 kOe) the final data point in magnetic relaxation follows the initial $M$($H$) curve, however, for $H$$>$10 kOe, final relaxation point does not follow the initial $M$($H$) curve. This is in stark contrast to the magnetic relaxation data observed for $H$$\parallel$$c$. Furthermore, inset (ii) of Fig. 4(a) shows the non-linear and a change of slope in-between (marked by arrows) behaviour of magnetic relaxation, which does not allow us to extract the relaxation rate for higher magnetic fields at temperatures above 6 K. Fig. 4(b) shows the magnetic field dependence of magnetic relaxation rate at different fixed temperatures for $H$$\parallel$$ab$.

\begin{figure}[htb]
\includegraphics[width=\linewidth]{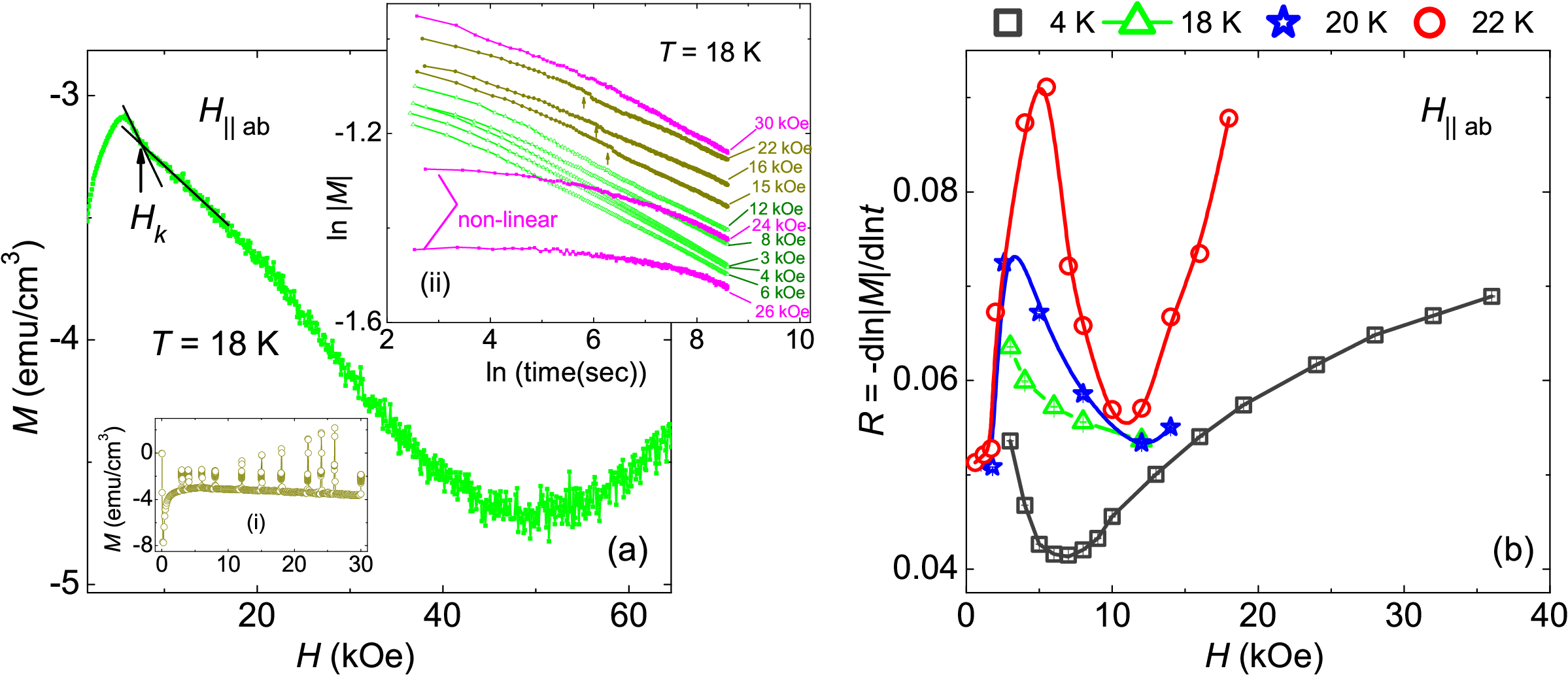}
\caption{(a) Initial branch of the isothermal magnetic field dependence of magnetization, $M$($H$), at 18 K. For magnetic fields just above $H_{on}$ a kink feature, $H_k$, is defined in $M$($H$). A pronounced noise in $M$($H$) is clearly observed as the magnetic field increases above $H_k$. Inset (i) shows the magnetic relaxation data measured at fixed fields on the initial branch of the $M$($H$) at 18 K. The field dependence of the magnetization after the relaxation changes drastically as compared to its initial branch. Inset (ii) shows ln$\lvert M \rvert$ vs. ln($t$) curves at different magnetic fields obtained from the magnetic relaxation data at 18 K. A non-linear behaviour as well as a step feature (marked by arrow) are observed in the ln$\lvert M \rvert$ vs. ln($t$) curves for $H$ $\geq$ 15 kOe. Therefore, the relaxation rate, $R$, could not be obtained at higher magnetic fields for $H$$\parallel$$ab$. (b) Relaxation rate, $R$, as a function of magnetic field at different temperatures.}
\label{fig:RH_ab}
\end{figure}

The temperature dependence of isofield relaxation rate for $H$$\parallel$$c$ and $H$$\parallel$$ab$ are shown in figures 5(a) and 5(b) respectively. For $H$$\parallel$$c$, the relaxation rate initially decreases with temperature, followed by a dip and then sharply increases at higher temperatures. Such sharp increase in relaxation rate at high temperatures may be ascribe to plastic vortex- creep. The position of the dip, $T_{cr}$ in each isofield $R$($T$) for $H$$\parallel$$c$, is well-matched with the one observed in isothermal $R$($H$), as observed in $H$-$T$ phase-diagram (see Fig. 6a), and has been attributed previously to the rhombic to square vortex lattice structural phase transition in superconductors \cite{Kopeliansky2010, Rosenstein2005, Sundar:2019a, Miu2020, Polichetti2021}. In the case of $H$$\parallel$$ab$ (Fig. 5(b)), the relaxation rate increases initially with temperature, followed by a slow reduction and a plateau, and subsequently followed by a sharp rise in relaxation rate. Unfortunately, a direct correlation between the SMP and these noticeable changes in $R$($T$) (Fig 5(b)) could not be established. 

To further understand the vortex-creep, we employed the collective creep theory \cite{Feigelman1989}, where the vortex pinning is assumed to be due to the weak randomly distributed defects (such as point-defects) in an elastic vortex system. In this theory, the variation of activation energy ($U$) with current density ($J$) can be described using the relation $U$ $\sim$ $U_c$($J_c$/$J$)$^{\mu}$. The glassy exponent, $\mu$$>$0, is temperature dependent and its value can inform us on the size of the vortex bundles experiencing creep. For a three-dimensional (3D) vortices, $\mu$ = 1/7, 3/2 or 5/2 and 7/9, have been predicted for the elastic vortex-creep of single vortices, small bundles, and large bundles of vortices respectively, in different ranges of current density. For 2D vortices, $\mu$ = 7/4, 13/16, and 1/2 is theoretically realized for elastic creep of small bundles, medium bundles, and large bundles of vortices, respectively. On the other hand, in the case of non-glassiness, $U$($J$) suppresses with $J$, the exponent $\mu$ is defined as $p$$<$0. In a recent theoretical development, $p$ = -0.75 is predicted for the plastic vortex-creep mediated by the mobility of edge dislocation through the vortex-density gradient \cite{Burlachkov2022}. It is to note that such plasticity is different than the one usually observed in atomic solids and is unique to the vortex state in a superconductor. In the Anderson-Kim theory \cite{Anderson1964, Lykov2014}, $p$ = -1 is predicted for thermally-activated vortex-creep. The values of $\mu$ and $p$ can be found by plotting the effective activation energy, $U_{eff}$ = $T/R$, with 1/$J$, which is given in the insets of Figs. 5(a) ($H$$\parallel$$c$) and 5(b) (for $H$$\parallel$$ab$). For $H$$\parallel$$c$, $\mu$ $\sim$ 1.5(1) is observed for magnetic field range of 3-20 kOe, which suggests the elastic creep of small bundles of 3D vortices, whereas, in the non-glassiness side, $\left|p\right|$ $\geq$ 1, for magnetic field range 3-10 kOe, suggest more than one mechanisms of vortex-creep. However, $p$ $\sim$ 0.6, indicates the density-gradient mechanism of plastic vortex-creep for 20 kOe \cite{Burlachkov2022}. For $H$$\parallel$$ab$ (Fig. 5(b)), two values of $\mu$ have been observed for each field in the different temperature ranges in $U_{eff}$ vs. 1/$J$ plot (slope in low-$T$ region for 5 kOe curve is not obtained due to insufficient data points). For 10 kOe, $\mu$ $\sim$ 0.64 and 0.87 is obtained in the temperature ranges of 8-14 K and 14-21 K respectively. These values seem compatible with the elastic vortex-creep of large bundles of 3D vortices. For 20 kOe, $\mu$ $\sim$ 0.56 and 1.34, were obtained in the temperature ranges 4-8 K and 10-18 K, respectively. This suggests a change in the size of the vortex bundles in elastic creep from large bundles at low temperature side to medium size bundles of 3D vortices at high temperatures. On the non-glassiness side, $p$ $\sim$ -0.58, -1, and -2.8 were obtained for 5 kOe, 10 kOe and 20 kOe respectively, where, except for 20 kOe, the $p$ values agree reasonably with the density-gradient mechanism of plastic vortex-creep \cite{Burlachkov2022}. However, large value of $p$ $\sim$ -2.8 for 20 kOe, does not support the existing mechanism of plastic vortex-creep, and therefore, it might be a representative of the entangled vortex phase at high magnetic fields that may likely have a different mechanism of vortex-creep. 

\begin{figure}[htb]
\includegraphics[width=\linewidth]{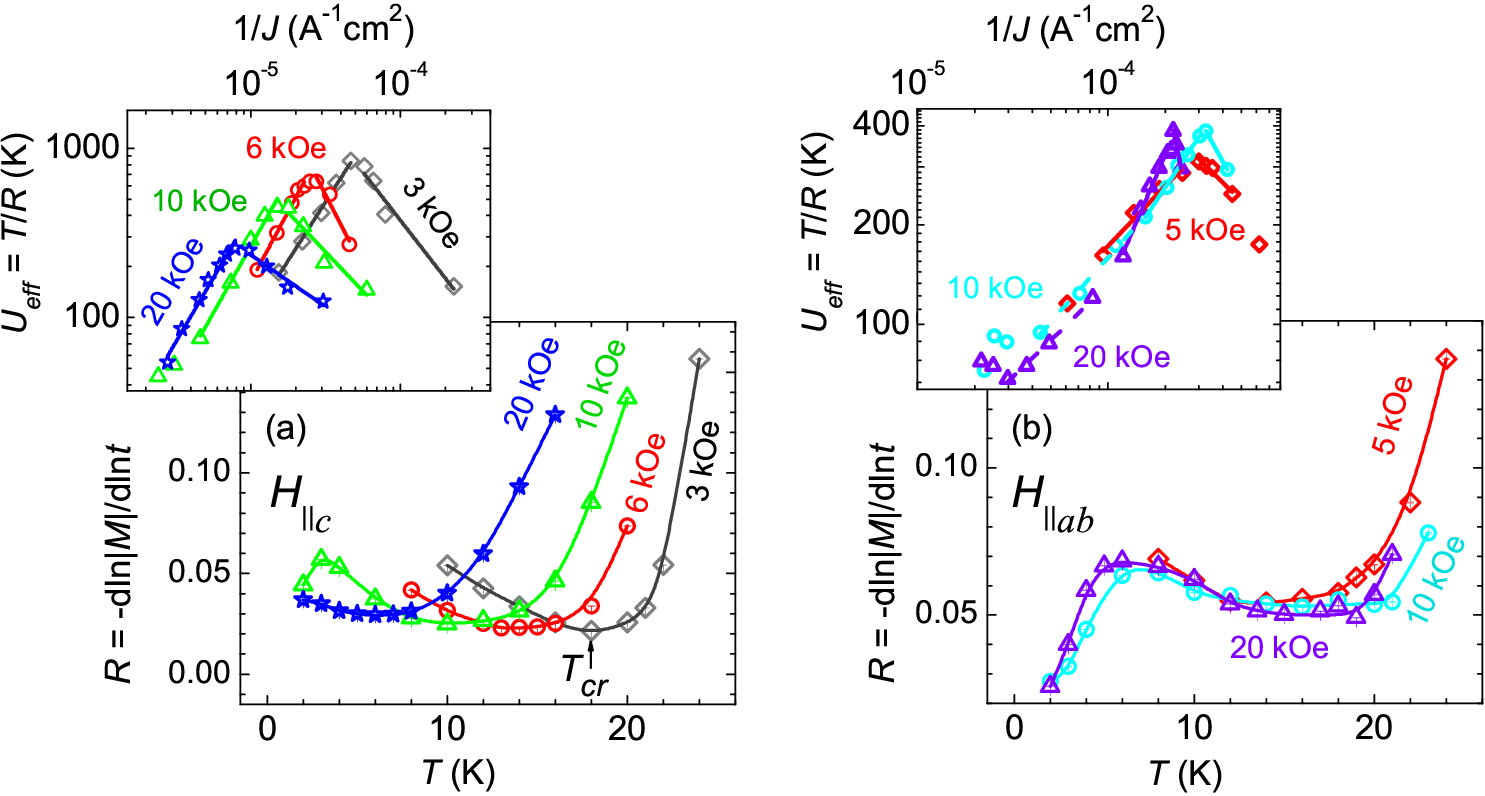}
\caption{(a) Isofield temperature variation of the relaxation rate, $R$($T$), at different magnetic fields for $H$$\parallel$$c$ (solid lines are guide to the eyes). Each $R$($T$) shows a dip at intermediate temperatures, defined as $T_{cr}$. Inset shows the activation energy, $U_{eff}$ = $T$/$R$, as a function of inverse of critical current density, 1/$J_c$ for $H$$\parallel$$c$. Solid lines are linear-fit to the data as explained in the main text. The observed exponents $\mu$ and $p$ across the peak in $U_{eff}$ vs. 1/$J_c$ suggest an elastic-to-plastic creep crossover. (b) Relaxation rate as a function of temperature, $R$($T$), at different fields for $H$$\parallel$$ab$ (solid lines are guide to the eyes). $U_{eff}$ vs. 1/$J_c$ plot for each $R$($T$) for $H$$\parallel$$ab$ is shown in the inset suggest an elastic-to-plastic creep crossover across peak.}
\label{fig:RT_both}
\end{figure}

Figure 6(a) shows the vortex-phase diagram for $H$$\parallel$$c$. It is interesting to observe that the $H_{on}$, $H_p$, $H_{cr}$ (stars) and $T_{cr}$ (pentagons) lines all show a strong concave shape, similar to other iron-pnictide superconductors \cite{Kopeliansky2010, Sundar2017a, Sundar:2019a}. The coinciding values of $T_{cr}$ and $H_{cr}$, defined as $H_{spt}$ in between the $H_{on}$ and $H_p$ lines, suggests a possible rhombic-to-square vortex lattice structural phase transition (SPT), initially identified for the La$_{2-x}$Sr$_x$CuO$_4$ by Rosenstein et al. \cite{Rosenstein2005} and later observed in 122-family of iron-pnictide superconductors \cite{Kopeliansky2010, Sundar2017a, Sundar:2019a}. As the SPT line is derived from the pinning mechanism associated to the $H_{on}$ line, its temperature dependence, in the present case, does not support the proposed theoretical models based on pinning due to the local variation in the $T_c$, $\delta$$T_c$-pinning, and the pinning due to the variation in charge carrier mean free path, $\delta$$l$-pinning, \cite{Blatter1994, Giller1997}. For a case of $\delta$$T_c$-pinning, $H_{on}$ saturates at low temperatures and show convex shape, whereas, for $\delta$$l$-pinning, $H_{on}$ increases with temperature \cite{Blatter1994, Giller1997, Giller1999}. It is important to note that these two models for $H_{on}$ neglected the effect of thermal fluctuations and are solely derived to capture the role of disorder. However, the $H_{on}$ in the present case can not be explained using either of the above mentioned models considering only disorder, and therefore thermal fluctuations must be playing an important role. Such thermal fluctuations lead to disrupt the symmetry of the rhombic vortex lattice state above $H_{on}$. As the field increases towards the SPT, the elastic squash modulus suppressed, leaving a stronger pinning and enhanced $J_c$, which renders a SMP in isothermal $M$($H$). Moreover, the strong concave-shape of the $H_{on}$, $H_p$ and SPT lines seem to be indicative of dominant role of thermal fluctuations \cite{Rosenstein2005}. As observed earlier, the pinning mechanism across $H_{on}$ may be understood as a dynamic crossover in the collective pinning, where a single vortex pinning below $H_{on}$ changes to a small bundle collective pinning above $H_{on}$ \cite{Sundar2017, SalemSugui2010, Abulafia1996}. The irreversibility line, $H_{irr}$($T$), is also plotted in the vortex phase-diagram in Fig. 6(a), $H_{irr}$ is defined as the magnetic field where $\Delta$$M$ = 0 (refer to Fig. S1 in supplementary information). It is observed previously that the rhombic-to-square vortex lattice SPT is usually associated with the elastic-plastic vortex creep across the $H_p$, which is already suggested in the inset of Fig. 5(a). In the latter part of the paper, a detailed analysis considering the scaling of the activation energy with magnetic field, using the Maley's analysis and collective creep theory, is also performed to show the temperature and magnetic field range for elastic (collective) and plastic vortex creep for both field orientations. 

\begin{figure}[htb]
\includegraphics[width=\linewidth]{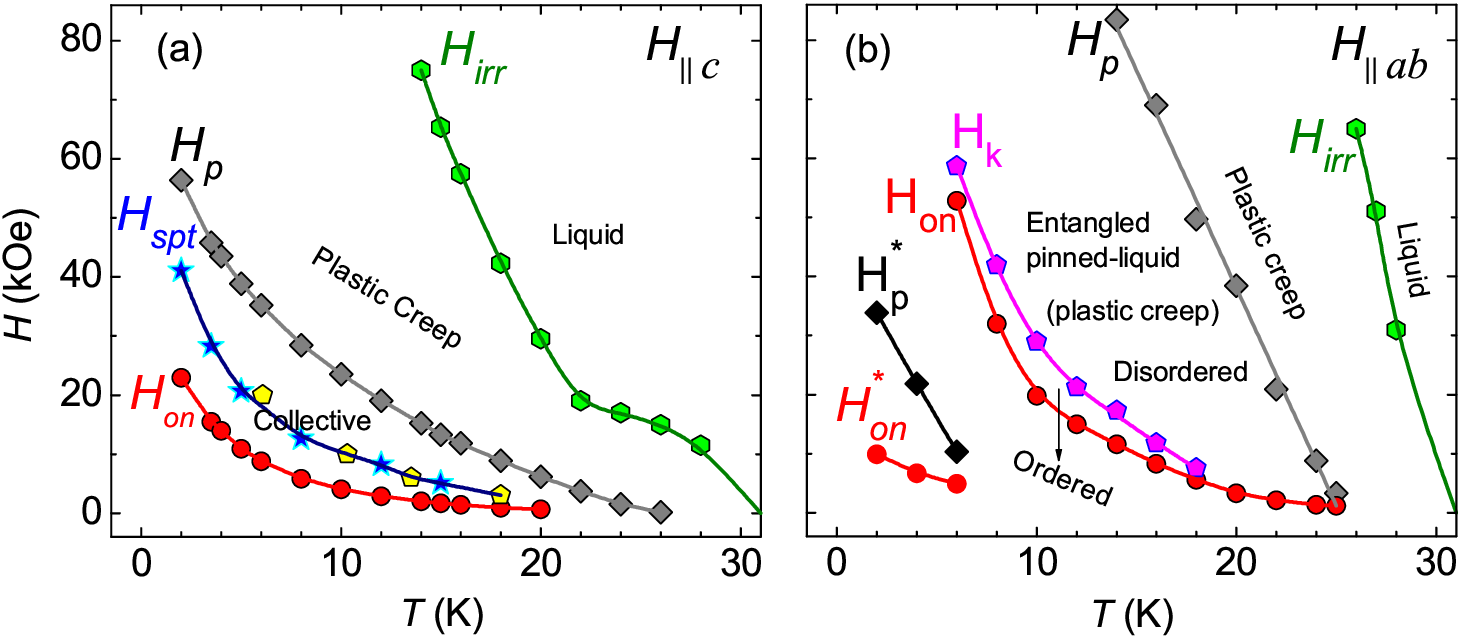}
\caption{Vortex phase-diagram, representing the temperature variation of different characteristic magnetic fields for (a) $H$$\parallel$$c$ (b) $H$$\parallel$$ab$. In panel (a) "star" and "pentagon" symbols in $H_{spt}$-line represent the $H_{cr}$ and $T_{cr}$ respectively as discussed in the main text. In both panels the solid lines are guides to the eyes.}
\label{fig:Phase-diagram}
\end{figure}
     
For $H$$\parallel$$ab$, the characteristic fields associated to the SMP at low-$T$ ($H^*_{p}$, $H^*_{on}$) and high-$T$ ($H_{p}$, $H_{on}$) and the kink field in the isothermal $M$($H$), $H_k$, and the irreversibility field, $H_{irr}$ are plotted in the vortex phase-diagram shown in Fig. 6(b). Clear signatures of the SMP and the associated magnetic field values are observed only for $T$ $\leq$ 25 K, whereas, the kink feature in $M$($H$) is found below 20 K. A distinct SMP (different than the high-$T$ SMP) at low-$T$ is only noticed for $T$ $\leq$ 6 K. Although, a SMP feature associated to the high-$T$ SMP is not observed at higher fields below 6 K, this might be either due to pronounced relaxation effects related to plastic creep or the SMP fields are beyond our maximum field limit. For high-$T$ SMP, the $H_k$ line marks the border above which a highly disordered entangled vortex phase is established, which leads to a SMP at higher fields. Such entangled disordered phase is usually characterized with plastic creep. It is interesting to note that the temperature dependence of $H_{on}$ and $H_k$ both show a strong concave-shape, which do not support the dislocation mediated plastic creep mechanism above $H_k$. However, the $H_p$ line does not show concave-shape instead it shows a slightly convex-shape. Therefore, it seems plausible that the mechanism of plastic creep below and above $H_p$ are different. Moreover, a density-gradient induced dislocation mediated plastic creep mechanism seems appropriate above $H_p$. 

A crucial question to be answered here is whether the disordered entangled phase above $H_k$ a solid-solid or solid-liquid phase transition. It is generally known that the melting line (transition to disordered vortex liquid phase) is observed in the reversible part of the isothermal $M$($H$), whereas, solid-solid (order-disorder) phase transition give rise to the SMP in isothermal $M$($H$). A disorder induced transition to an entangled solid vortex phase (order-disorder transition) has been observed in cuprate superconductors, such as Nd$_{1.85}$Ce$_{0.15}$CuO$_{4-\delta}$ (NCCO) \cite{Giller1997}, YBa$_2$Cu$_3$O$_{7-\delta}$ (YBCO) \cite{Giller1999}, and BSCCO \cite{Khaykovich1996}. However, the temperature dependence of the transition line ($H_k$) is quite different in each superconductor, $H_k$ decreases monotonically with temperature and shows a convex shape for NCCO, whereas it shows a non-monotonic temperature variation for YBCO and temperature independent behaviour for BSCCO. No such temperature dependence of $H_k$ matches with the one observed in this work. Therefore, we may consider that this transition is not a pure solid-solid phase transition. Another aspect is that a pure solid-solid phase transition is driven by a disorder-induced fluctuation, which shows a monotonic suppression of $H_k$ (convex shape) and an increasing behaviour of $H_k$ with $T$ for $\delta$$T_c$-pinning and $\delta$$l$-pinning respectively \cite{Giller1999}. In the present case, the $H_{k}$ line shows a strong suppression with increasing $T$ with a concave-shape, which suggests that the order-disorder transition is due to the dominant role of thermal fluctuations. A unified picture of order-disorder vortex phase transition has been developed by Radzyner et al. \cite{Radzyner2002a, Radzyner2002}. In this picture, the relative contributions of disorder and thermally induced fluctuations may lead to the disordered entangled solid state, pinned-liquid as well as liquid vortex state, where, the melting transition line (liquid vortex state), considers the vanishing pinning energy, shows a concave shape, and the entangled solid transition line neglects the role of thermal energy (convex shape). The order-disorder transition in the present case is observed in the irreversible region of the isothermal $M$($H$) and, unlike the melting transition, shows a strong temperature dependence with a concave-shape. Therefore, it resembles an intermediate situation of entangled pinned-liquid vortex state, where both thermal energy and pinning energy are contributing together. As the concave shape of the $H_k$ line is close to the one observed for the melting line by Radzyner et al., we conclude that the thermal energy has a dominant contribution for order-disorder transition in the present case. This explains the pronounced noise observed in the isothermal $M$($H$) above $H_k$. A similar temperature dependence of $H_k$ (order-disorder transition) is also found in La$_{2-x}$Sr$_x$CuO$_4$ and has been explained as an ordered to disordered pinned-liquid vortex state transition \cite{Radzyner2002}. The characteristic fields associated to the low-$T$ SMP ($H^*_{p}$, $H^*_{on}$) are also plotted in the vortex phase-diagram shown in Fig. 6(b). Since, low-$T$ SMP is observed for $T$ $\leq$ 6 K, we do not have a wide enough temperature range to explore the temperature dependence of $H^*_{p}$, $H^*_{on}$. Moreover, the shape of the low-$T$ SMP in isothermal $M$($H$) is quite different than the SMP at high-$T$. This might be due to the suppressed thermal fluctuations at low temperatures, which changes the relative contribution of pinning energy, elastic energy and thermal energy at low temperature and alters the vortex dynamics. 

\begin{figure}[htb]
\includegraphics[width=\linewidth]{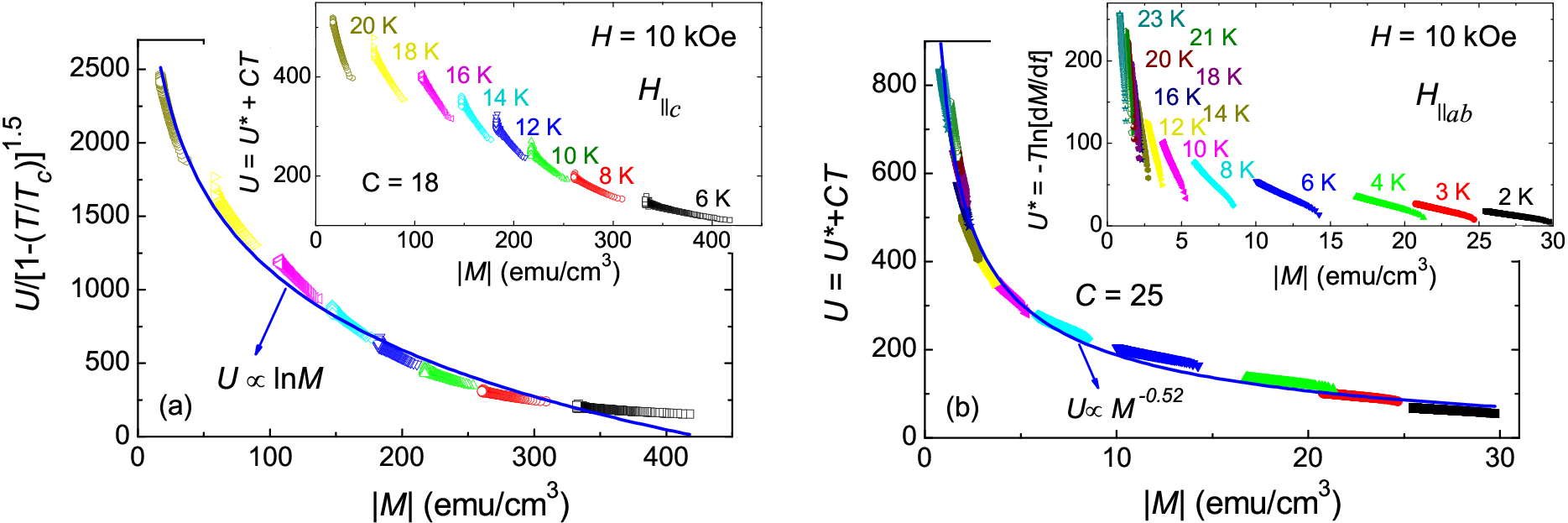}
\caption{(a) Scaling of activation energy curves, $U(M)$=-$T$dln$M(t)$/d$t$ + $C$$T$ = $U^* + CT$, using a function $g$($T$)=(1-$T$/$T_c$)$^{3/2}$, for $H$$\parallel$$c$. Solid line shows a $\sim$ln$M$ behaviour of the scaled curve. Inset shows the $U$($M$) curves, at $H$ = 10 kOe for $T$ = 6-20 K, without scaling. (b) For $H$$\parallel$$ab$, scaling of activation energy curves, $U$($M$), is achieved without using the function $g$($T$). Scaled $U$($M$) curves for $H$$\parallel$$ab$ follow $\sim$ $M^{-0.52}$ variation (solid line). Inset shows $U^*$($M$) curves for $T$ = 2-23 K, and $H$ = 10 kOe.}
\label{fig:Maley_both}
\end{figure}

\begin{figure}[htb]
\includegraphics[width=\linewidth]{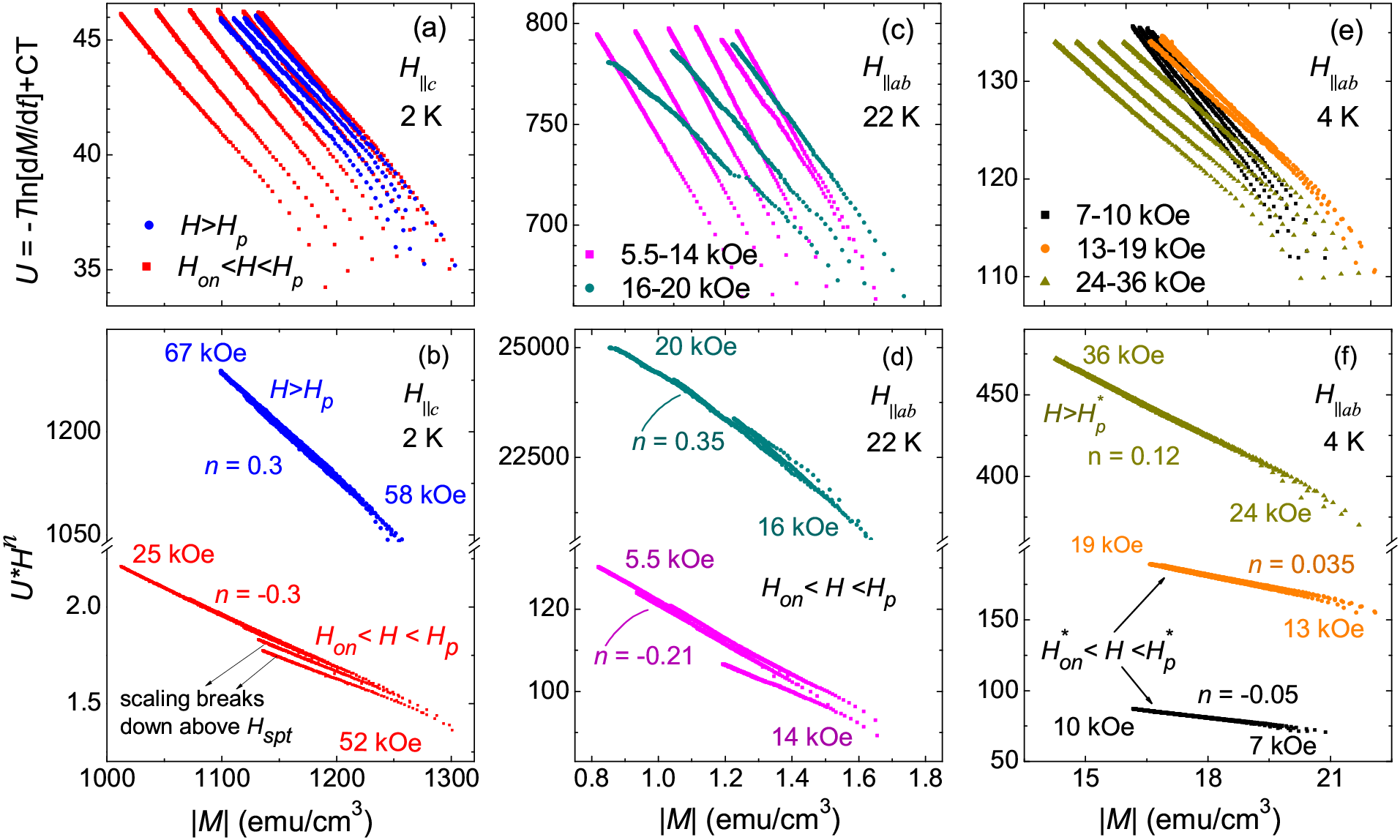}
%\begin{center}
%\includegraphics[scale=0.45]{Fig8.eps}
%\end{center}
\caption{(a) Activation energy as a function of magnetization, $U$($M$), obtained for magnetic fields across the SMP at 2 K for $H$$\parallel$$c$. (b) Scaling of $U$($M$) curves, presented in panel (a), with $H^n$, suggest an elastic-to-plastic creep crossover across SMP. In the elastic creep regime, scaling breaks down for $H$ above a magnetic field associated to the vortex-lattice structural phase transition, $H_{spt}$. (c) $U$($M$) curves obtained at 22 K for different magnetic fields $H$$<$$H_p$ for $H$$\parallel$$ab$. (d) Scaling of $U$($M$) curves in panel (c) with $H^n$ shows that an elastic creep changes to plastic creep well before approaching $H_p$. (e) $U$($M$) curves for different magnetic fields across the SMP at 2 K for $H$$\parallel$$ab$. (f) Scaling of $U$($M$) curves in panel (e) shows an elastic-to-plastic creep crossover for $H^*_{on}$$<$$H$$<$$H^*_{p}$. A plastic creep is also observed for $H$$>$$H^*_{p}$ but with different value of exponent $n$. This suggests that the nature of the plastic pinning observed below and above $H^*_{p}$ are distinct.}
\label{fig:EP-scaling}
\end{figure}

To further explore the vortex dynamics associated to the SMPs, we employed the activation energy ($U$) obtained from the magnetic relaxation data, $M$($t$), using a criterion developed by Maley et al \cite{Maley1990}, which yield, $U(M)$=-$T$dln$M(t)$/d$t$ + $C$$T$ = $U^* + CT$, where $C$ is a constant which depends on the vortex hopping distance, attempt frequency, and the sample size. For $H$$\parallel$$c$, the inset of the Fig. 7(a) shows the $U$($M$) plot for $C$ = 18, for magnetic relaxations measured at 10 kOe for different temperatures in the range 6-20 K. It can be observed that the $U(M)$ (inset of Fig. 7(a)) does not show a smooth behaviour for $C$ = 18, in the whole temperature range of measurements. However, an appropriate value $C$ not always leads to a smooth function of $U$($M$). A smooth variation of $U(M)$ can be achieved by scaling the $U(M)$ with the function, $g$($T$)=(1-$T$/$T_c$)$^{3/2}$, that correlates the coherence length with temperature \cite{McHenry1991}. The variation of $U$/$g(T$) vs. $\left|M\right|$ follows a ln$M$ dependence, as shown in the main panel of Fig. 7(a). For $H$$\parallel$$ab$, a power law dependence of $U$($M$), in the temperature range 2-23 K, is obtained for $C$ = 25, and unlike for $H$$\parallel$$c$, it does not require a scaling with a function $g$($T$), as shown in Fig. 7(b). A non-smooth variation of $U^*$ = -$T$dln$M(t)$/d$t$ as a function of $\left|M\right|$ is shown in the inset of Fig. 7(b). The obtained values of the parameter $C$ = 18 ($H$$\parallel$$c$) and 25 ($H$$\parallel$$ab$), are used to estimate the activation energy, $U(M)$=-$T$dln$M(t)$/d$t$ + $C$$T$, using the magnetic relaxation data measured across the SMP in isothermal $M$($H$). The investigation of the activation energy, $U(M)$, using collective creep theory, will assist us to understand the vortex dynamics in the different magnetic field regimes above and below the SMP.   

\begin{figure}[htb]
\includegraphics[width=\linewidth]{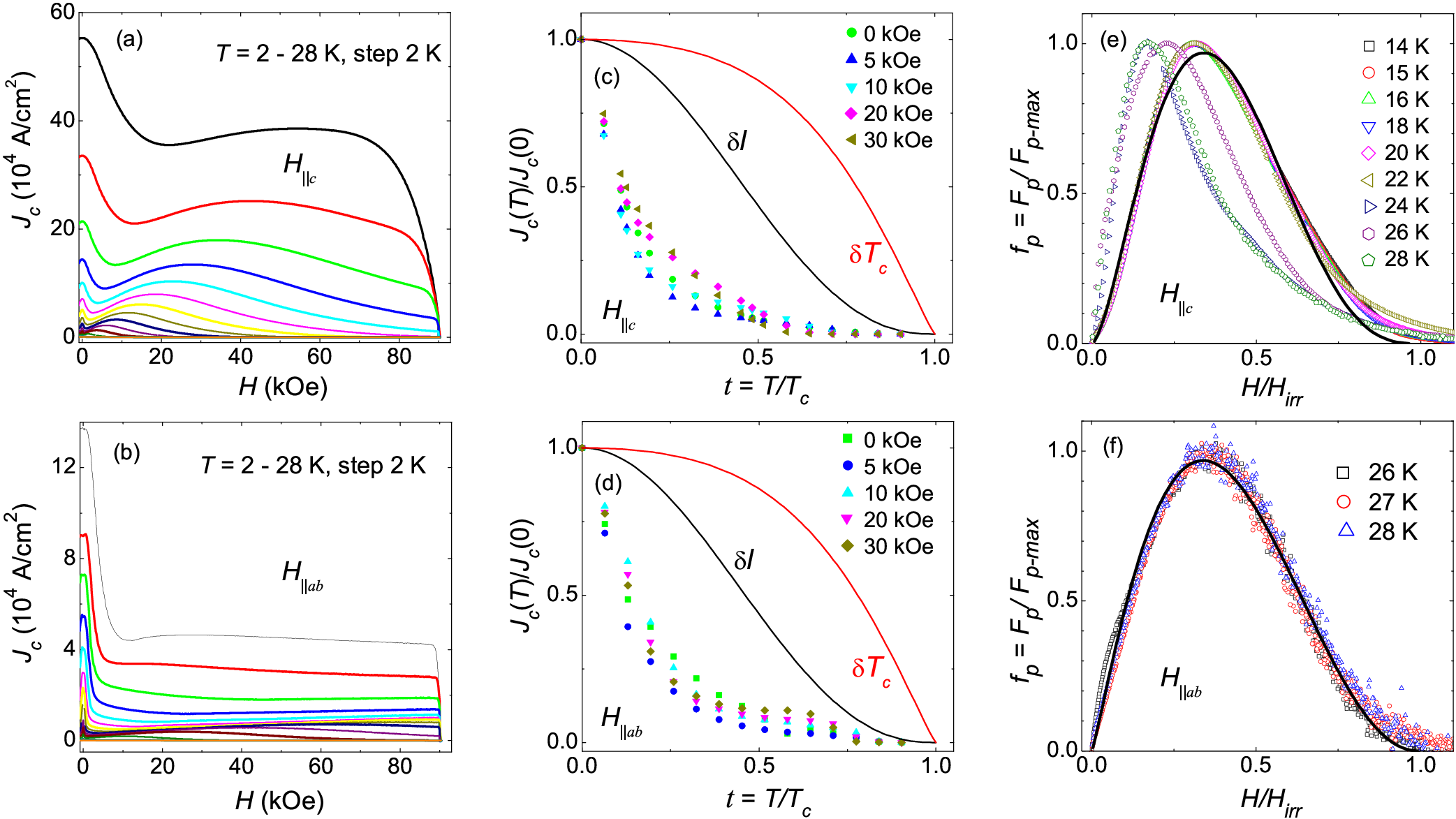}
%\begin{center}
%\includegraphics[scale=0.45]{Fig9.eps}
%\end{center}
\caption{Magnetic field dependence of critical current density at different fixed temperatures for (a) $H$$\parallel$$c$, and (b) $H$$\parallel$$ab$. (c) Critical current density normalized with its zero temperature limit, $J_c$($T$)/$J_c$(0), as a function of reduced temperature, $t$ = $T$/$T_c$, for various fields. Solid lines show the $J_c$($T$)/$J_c$(0) vs. $t$ curves for $\delta$$l$, and $\delta$$T_c$ pinning models discussed in the main text. (e, f) Normalized pinning force density, $F_p$/$F_{p-max}$, as a function of reduced magnetic field, $h$ (= $H$/$H_{irr}$) at different temperatures below $T_c$ for $H$$\parallel$$c$, and $H$$\parallel$$ab$ respectively. Scaled $F_p$/$F_{p-max}$ curves show a peak close to $h$ = 0.33 for both crystal orientations, which indicate a vortex-pinning due to point disorder. For $H$$\parallel$$c$ (panel e), the peak shifts to lower $h$, indicating more than one type of pinning sites involve in vortex pinning near $T_c$. For both panels (e, f), the scaled $F_p$/$F_{p-max}$ curves are fitted to an expression, $F_p$/$F_{p-max}$ = $A h^p (1-h)^q$, where the peak position, $h_{max}$ = $p$/($p$+$q$), is found to be very close to the peak in scaled pinning force curves.}
\label{fig:Jc}
\end{figure}  

According to collective creep theory, the activation energy can be expressed as $U$($B$,$J$) = $B^{\nu}$$J^{-\mu}$ $\approx$ $H^{\nu}$$M^{-\mu}$, where the exponents $\nu$ and $\mu$ depend on the specific flux-creep \cite{Feigelman1989, Abulafia1996}. It is established that for collective creep (elastic) the activation energy increases with magnetic field and when it shows a decreasing behaviour with magnetic field, it is explained in terms of plastic pinning \cite{Abulafia1996}. Therefore, a positive value of exponent $\nu$ represents a collective (elastic) creep and a negative value of $\nu$ manifests a plastic creep. Such creep crossover has been observed in several other studies on vortex dynamics in high-$T_c$ superconductors \cite{Sundar:2019a, Said:2020, Llovo:2021, Sundar2017, SalemSugui2010}. Fig. 8(a) shows the $U$($M$) curves, obtained using Maley's criterion,  at $T$ = 2 K for magnetic fields $H$$>$$H_p$ and $H_{on}$$<$$H$$<$$H_p$. A slight change in the slope of the $U$($M$) curves for magnetic fields above and below $H_p$ suggests a change in creep mechanism. For a better insight, a scaling of $U$($M$) with $H^n$ is obtained. Fig. 8(b) depicts the $U$*$H^n$ scaling as a function of $\left|M\right|$, where a separate scaling for field above and below $H_p$ is observed. A negative $n$ = -0.3, for $H_{on}$$<$$H$$<$$H_p$ evidenced the collective creep (elastic) and a positive $n$ = 0.3, indicates the plastic creep for $H$$>$$H_p$. The absolute value of the exponent, other than $n$ = 0.75, suggests that the plastic creep mechanism in the present case is different than the density-gradient induced dislocation mediated plastic creep. However, a crossover from elastic-to-plastic creep explains the origin of SMP for $H$$\parallel$$c$. Moreover, it is also observed that for $H_{on}$$<$$H$$<$$H_p$, the scaling breaks down near the field related to the structural phase transition ($H_{spt}$). This signifies that the pinning energy landscape changes after the vortex lattice changes from rhombic-to-square vortex lattice above $H_{spt}$ (or $H_{cr}$). 

For $H$$\parallel$$ab$, to explore the vortex dynamics at temperatures where no kink feature ($H_k$) is observed in the isothermal $M$($H$) related to the order-disorder transition, we plotted the $U$($M$) curves for 22 K, and $H$ = 5.5 - 20 kOe ($H_{on}$$<$$H$$<$$H_p$) in Fig. 8(c). We noticed that $U$($M$) curves for $H$ = 5.5 - 14 kOe and $H$ = 16-20 kOe, show quite distinct slopes. Scaling of $U$($M$) with $H^n$, in Fig. 8(d), shows an elastic-to-plastic creep crossover well before approaching $H_p$. It is further noticed that the plastic creep in the present case develops for fields where the noise in the isothermal $M$($H$) becomes significant. The plasticity of the vortex creep well below $H_p$ seems to be related to the entangled pinned-liquid which is expected above the order-disorder transition for $H$$\parallel$$ab$. Furthermore, we realized that the order-disorder transition at $H_k$ appears to be due to the intricate cooperative role of pinning energy, elastic energy and thermal energy, where the dominant proportion of thermal energy as compared to the elastic energy and pinning energy at higher temperatures suppressed the kink feature. 

To understand the vortex dynamics associated to the SMP observed at low temperatures for $H$$\parallel$$ab$, in Fig. 8(e), we plotted $U$($M$) at 4 K for magnetic fields above and below $H^*_p$. Interestingly, it is observed that the $U$($M$) curves for $H^*_{on}$$<$$H$$<$$H^*_p$ show a different behaviour for magnetic field ranges 7-10 kOe and 13-19 kOe. Moreover, the $U$($M$) curves for $H$$>$$H^*_p$ have distinctly different slope compared to the one for $H$$<$$H^*_p$. The scaling of $U$($M$) curves with $H^n$ show an elastic-to-plastic type crossover at an intermediate field between $H^*_{on}$ and $H^*_p$. Although the negative value of $n$ is indicative of plastic creep, its absolute value is quite smaller than the one proposed for dislocation mediated plastic creep ($n$ = 0.75). Furthermore, the $n$ value for collective creep does not match the literature for small, medium and large size vortex bundles. Magnetic fields above $H^*_p$ yield $n$ = 0.12, which suggests the plastic vortex creep. It is interesting to note that the dislocation mediated plastic vortex creep does not explain the plasticity in the this sample for both crystal orientations, and its origin must be further investigated. 

The field dependence of the critical current density, $J_c$($H$), at different temperatures, for $H$$\parallel$$c$ and $H$$\parallel$$ab$, is shown in Fig. 9(a) and 9 (b) respectively. $J_c$ was estimated from the isothermal $M$($H$) through the Bean's critical-state model \cite{Bean1964}, using the expression, $J_c$ = 20$\Delta$$M$/$a$(1-$a$/3$b$), where $b$$>$$a$, are the crystal dimensions of the plane perpendicular to $H$, and $\Delta$$M$(emu/cm$^3$) is the difference between the magnetic field decreasing and increasing branch of the isothermal $M$($H$) curves \cite{Umezawa1987}. The $J_c$ values for $H$$\parallel$$c$ are an order of magnitude larger than $J_c$ for $H$$\parallel$$ab$. Unlike for $H$$\parallel$$c$, the $J_c$ for $H$$\parallel$$ab$ at low temperatures suppressed drastically with magnetic field and shows a near-plateau behaviour at higher magnetic fields. Although the $J_c$ for $H$$\parallel$$c$ at $T$$<$10 K is $\sim$ 10$^5$ A/cm$^2$, meeting the requirements for high-magnetic-field applications in superconducting materials, the use of expensive platinum (Pt) in this compound limits its practicality for fabricating superconducting wires.

The temperature dependence of $J_c$ normalized with its zero temperature limit, $J_c$($T$)/$J_c$(0), can be useful to access the $\delta$$l$-pinning or a $\delta$$T_c$-pinning. The temperature dependence for $\delta$$l$-pinning is, $J_c(T)$/$J_c(0)$ = $(1+t^2)$$^{-1/2}$$(1-t^2)$$^{5/2}$, and, for $\delta$$T_c$-pinning is, $J_c(T)$/$J_c(0)$ = $(1-t^2)$$^{7/6}$$(1+t^2)$$^{5/6}$, where $t$ = $T$/$T_c$ \cite{Griessen1994}. Figure, 9(c) and 9(d) show the $J_c$($T$)/$J_c$(0) vs $t$ curves obtained for different field values including at zero-field. For both crystal orientations, $J_c$($T$)/$J_c$(0) follow a very similar temperature dependence with a strong concave shape as also seen in vortex-phase diagram in Fig. 6 (a, b). Moreover, comparing the $t$ dependence of $J_c$($T$)/$J_c$(0) with the one proposed for $\delta$$l$-pinning and $\delta$$T_c$-pinning models, we can see that it does not follow either of the pinning models. Moreover, it is interesting to note that the behavior of $J_c$($t$) closely resembles the qualitative pattern observed for characteristic fields in the $H$-$T$ phase diagrams shown in Fig. 6. This observation hints at the potential influence of thermal fluctuations on $J_c$, even at low temperatures. Additionally, we found that the temperature variation of $J_c$ can be reasonably fitted by employing both weak and strong pinning models \cite{Galluzzi2019}. Specifically, the data at low fields follow the weak pinning model, while at higher fields, a combination of weak and strong pinning explains the data better (refer to Fig. S2 and Fig. S3 in the supplementary information for more details). 

To explore the disorder-type responsible for vortex pinning, we estimated the pinning force density using, $F_p$ = $J_c$$\times$$H$, where $J_c$ is the critical current density and $H$ is the magnetic field. The normalized pinning force density, $F_p$/$F_{p-max}$ as a function of reduced magnetic field, $h$ (= $H$/$H_{irr}$), is plotted in Figs. 9(e) and (f) for $H$$\parallel$$c$ and $H$$\parallel$$ab$ respectively for different temperatures. For $H$$\parallel$$c$, the $F_p$/$F_{p-max}$ curves below 24 K show good scaling with $h_{max}$ $\sim$ 0.32, where $h_{max}$ is the reduced field related to $F_p$/$F_{p-max}$ = 1. However, at higher temperatures, close to $T_c$, $F_p$/$F_{p-max}$ curves show poor scaling and the peak shifts to lower $h$ values. This might be due to the sample inhomogeneity which affects the pinning landscape near $T_c$. A good scaling of $F_p$/$F_{p-max}$ vs. $h$ curves with $h_{max}$ $\sim$ 0.34, is obtained for $H$$\parallel$$ab$ in Fig. 9(f). The collapse of the $F_p$/$F_{p-ma}$ vs. $h$ curves, obtained at different temperatures, into a single curve can be utilized to find the dominant source of pinning using the well-known Dew-Hughes model \cite{DewHughes1974}. In this model, the normalized pinning force density can be expressed as, $F_p$/$F_{p-max}$ = $A h^p (1-h)^q$, where the parameters $p$ and $q$ are related to the type of disorder through $h_{max}$ = $p$/($p$+$q$) \cite{DewHughes1974}. For $H$$\parallel$$c$, the fitting of the scaled curves using the expression given above provides, $p$ = 1.62, $q$ = 3.13 and therefore, $h_{max}$ = $p$/($p$+$q$) = 0.34, which is very close to the one obtained from scaling of the experimental data. For pinning due to point disorder, the Dew-Hughes proposed, $p$ = 1, $q$ = 2, and $h_{max}$ = 0.33. Although the fitted values of $p$ and $q$ are different than the expected ones, the maximum in $F_p$/$F_{p-max}$ is observed at 0.32, which suggests that the point disorder is the dominant source of vortex pinning for $T$$<$24 K. However, the maximum in $F_p$/$F_{p-max}$ shifts to lower $h$ at higher temperatures, which indicates that more than one pinning centers are involved in vortex pinning near $T_c$ for $H$$\parallel$$c$. For $H$$\parallel$$ab$ in Fig. 9 (f), the similar fitting of scaled curves using Dew-Hughes's expression yields, $p$ = 1.17, $q$ = 2.3, and therefore, $h_{max}$ = $p$/($p$+$q$) = 0.34. The value of $h_{max}$ matches very well with the one found in the experimental curves. Therefore, we conclude that point disorder (weak pinning centers) is the dominant source of vortex pinning for both crystal orientations. 

\section*{Conclusion}

In conclusion, we reported a detailed investigation of vortex dynamics and second magnetization peak (SMP) in a triclinic iron pnictide superconductor (Ca$_{0.85}$La$_{0.15}$)$_{10}$(Pt$_3$As$_8$)(Fe$_2$As$_2$)$_5$, with $T_c$ $\sim$ 31 K, via dc magnetization measurements for $H$$\parallel$$c$ and $H$$\parallel$$ab$. A SMP feature, although observed for both crystal orientations, show a different behaviour and therefore, distinct reasons for its appearance in isothermal $M$($H$) below $T_c$. For $H$$\parallel$$c$, the SMP is similar to previously reported in other iron pnictides, and it can be well explained with an elastic-to-plastic creep crossover across the SMP. In addition, a possible rhombic-to-square vortex lattice phase transition is also observed for fields in between the onset-field and the peak-field of SMP. On the other hand, for $H$$\parallel$$ab$, the SMP at low-$T$ and high-$T$ regions below $T_c$ is found to be of a distinct nature. For $T$$>$6 K, an order-disorder vortex phase transition is found, which shows an entangled pinned vortex-liquid in the disordered vortex phase. In addition, a noisy $M$($H$) data is also observed in the disordered vortex phase above the transition. For $T$$<$6 K, an elastic-to-plastic creep is observed for fields just below the SMP and that persists for field above SMP as well. Moreover, the variation of the critical current density with temperature does not follow the $\delta$$l$ and $\delta$$T_c$ vortex pinning models for both crystal orientations. Furthermore, through an analysis using the pinning force density curves, it is found that the point defects are the main source of vortex pinning for both crystal geometries. Vortex phase-diagrams for $H$$\parallel$$c$ and $H$$\parallel$$ab$ are presented showing the characteristic temperatures and magnetic fields in this unique iron-pnictide superconductor.

\section*{Methods}

Single crystal of La-10-3-8 was grown using the self-flux method \cite{Seo2017, Ni2013, Choi2020, Xiang2012, Pan2017}.  Magnetization measurements were performed on a flat-plate like crystal of mass $\sim$ 7 mg, and dimensions $\sim$2$\times$2.5$\times$0.39 mm$^3$, using a commercial vibrating sample magnetometer (VSM), from Quantum-Design, USA. The single crystal was mounted on a quartz sample holder for measurement with magnetic field parallel to the ab-plane ($H$$\parallel$$ab$), whereas, for magnetic field perpendicular to the plane ($H$$\parallel$$c$), the crystal was held between the two quartz cylinders in a brass sample holder. Apiezon N grease was used as an adhesive for sample mounting. Temperature dependence of magnetization was measured in zero-field-cooled (ZFC) as well as in field-cooled (FC) modes. In ZFC mode, a desired low temperature below $T_c$ was achieved in the absence of magnetic field, then a magnetic field was applied, and the measurement was performed while warming. This is followed by measuring while cooling with the magnetic field still on (FC). Five quadrant magnetic hysteresis loops, isothermal $M(H)$, in the superconducting state were measured with a ZFC mode. Magnetization as a function of time, $M$($t$), was measured at various fixed temperatures and magnetic fields below $T_c$, with a span time of $\sim$90 minutes.

%\bibliographystyle{}
%\bibstyle{}
%\bibliography{Ref_10-3-8}

\section*{Acknowledgements}  
PVL is supported by an M.Sc. grant from Conselho Nacional de Desenvolvimento Científco e Tecnológico (CNPq). LG was supported by Fundação Carlos Chagas Filho de Amparo à Pesquisa do Estado do Rio de Janeiro (FAPERJ), Projects E-26/010.001497/2019, E-26/211.270/2021, and E-26/200.964/2022, and CNPq, project 308899/2021-0. This work is also supported by the National Key Research and Development Program of China (Grant No. 2018YFA0704200), the Strategic Priority Research Program (B) of the CAS (Grants Nos. XDB25000000), the National Natural Science Foundation of China (Grant Nos. 11822411 and 11961160699), and the Youth Innovation Promotion Association of CAS (Grant No. Y202001). W.-S. H. was supported by  the Postdoctoral Innovative Talent program (Grant No. BX2021018), and the China Postdoctoral Science Foundation (Grant No. 2021M700250).

\section*{Author contributions statement}

ZZL, HL, WH, and SLL prepared the single crystal. SS and S-S conceived the experiment. PVL, SS and LG conducted the magnetization measurements, SS and PVL analyzed the data. SS wrote the manuscript. All authors reviewed the manuscript. 

\section*{Additional information}
The datasets used and/or analysed during the current study available from the corresponding author on reasonable request.
Correspondence and materials should be addressed to SS.

\textbf{Competing interests} 
The authors declare no competing interests.

\end{document}